\documentclass{article}
\usepackage{spconf,amsmath,graphicx}
\usepackage{xcolor}
\usepackage{pbox}
\usepackage{hyperref}
\usepackage{url}
\usepackage{adjustbox}

\newcommand\glpair[2]{{\small\textsf{#1}}-{\scriptsize\textsf{#2}}}
\newcommand\gl[1]{{\small\textsf{#1}}}
\newcommand{\matr}[1]{\mathbf{#1}} 

\title{SCORE DIFFICULTY ANALYSIS FOR PIANO PERFORMANCE EDUCATION \\BASED ON FINGERING}
\name{Pedro Ramoneda\sthanks{Corresponding author: \url{pedro.ramoneda@upf.edu}}, Nazif Can Tamer, Vsevolod Eremenko, Xavier Serra, Marius Miron}
\address{Music Technology Group, Universitat Pompeu Fabra, Barcelona}

\begin{document}
\ninept
\maketitle
\begin{abstract}

\noindent In this paper, we introduce score difficulty classification as a sub-task of music information retrieval (MIR), which may be used in music education technologies, for personalised curriculum generation, and score retrieval. We introduce a novel dataset for our task, Mikrokosmos-difficulty, containing 147 piano pieces in symbolic representation and the corresponding difficulty labels derived by its composer Béla Bartók and the publishers. As part of our methodology, we propose piano technique feature representations based on different piano fingering algorithms. We use these features as input for two classifiers: a Gated Recurrent Unit neural network (GRU) with attention mechanism and gradient-boosted trees trained on score segments. We show that for our dataset fingering based features perform better than a simple baseline considering solely the notes in the score. Furthermore, the GRU with attention mechanism classifier surpasses the gradient-boosted trees. Our proposed models are interpretable and are capable of generating difficulty feedback both locally, on short term segments, and globally, for whole pieces. Code, datasets, models, and an online demo are made available for reproducibility. 

\end{abstract}


\section{Introduction}
\label{sec:intro}

Classification of music corpora is a problem well-studied under Music Information Retrieval (MIR), which is often targeted from the listeners' perspective as exemplified in genre/style~\cite{ghosal2018music,weiss2015tonal} and emotion~\cite{fukayama2016music,yang2010ranking} classification. In a paradigm shift, music may be classified from the point of view of the performer by focusing on the required performance skills~\cite{chiu2012study,RamonedaTFM}, which is a newly emerging field of study. This paper focuses on music classification of performance difficulty, with  applications in the formation of large pedagogical score databases, personalised score recommendation systems, and as an aid to both individual instrument learners and music teachers. Towards helping the students in determining where to focus their effort, and thus, increasing the efficacy in self-induced music studies, we aim at giving feedback on relative local difficulty over multiple segments of a piece.

The difficulty, by definition, is a subjective measure with multiple dimensions (i.e. expressivity, rhythm, sound, and technique~\cite{neuhaus2008art}). Publishers, examination boards, and online repositories classify scores based on performance difficulty. Many exam boards relate these rankings with the school grades. A widely known publisher, Henle, releases piano difficulty rankings for their scores in the range 1-9, but the range is more tailored towards professionals. Certain instrument forums and websites such as 8note also release difficulty grades in different ranges, generally three or four grades from beginner to advanced. When it comes to the difficulty datasets in MIR, both previous datasets~\cite{sebastien2012score,chiu2012study} are not publicly available.

As a first contribution, we release Mikrokosmos-difficulty~\footnote{Dataset available at:\\\url{https://github.com/PRamoneda/Mikrokosmos-difficulty}\\DOI: \url{10.5281/zenodo.6092709}} --  a benchmark dataset for piano score difficulty analysis derived from a corpus of 147 educational pieces authored by Béla Bartók for use in piano education. To our best knowledge, this is the first open dataset containing piano scores ranked in terms of difficulty and matching different technical levels where all the scores are composed by a single composer. 
%
%
Alongside the data, we also provide three different difficulty labels to study the subjectivity of performance difficulty: the first labels are the order of pieces generated by the composer himself, the second is the book divisions by the original publisher, and the third is difficulty labels in the range 1-9 by the publisher Henle, respectively. Since all the scores are composed by the same composer, the difficulty rankings are less prone to style biases, and more focused on technique difficulty.

As a second contribution, we introduce several piano technique features and two classification algorithms capable of giving both score-level and segment-level difficulty feedback~\footnote{Code and models available at:\\  \url{https://github.com/PRamoneda/ICASSP22}}, whilst being trained solely using score-level labels. To that end, we model the score with a novel feature representation based on piano fingering, and taking that as the input, we propose two difficulty classification methods: (i) gradient-boosted trees~\cite{chen2016xgboost} applied to short-term segments and (ii) GRU neural network~\cite{maghoumi2019deepgru} with an attention mechanism. We want the selected classifiers to give feedback related to the local difficulty of a piece allowing students and teachers to focus and improve on the most difficult passages. Further, we provide a corpus visualisation tool for exploring local difficulty representations derived from both attention weights and the segment-level classification models. The remainder of this paper is organized as follows: in Section~\ref{sec:relatedwork}, we give the previous related work, in Section~\ref{sec:dataset} we introduce the Mikrokosmos-difficulty dataset, in Section~\ref{sec:methods} we propose our difficulty analysis methods and give the results of our experiments in Section~\ref{sec:experiments}.

\section{Relation with previous work}
\label{sec:relatedwork}
In this section we refer to the main methods to model difficulty in piano repertoire~\cite{sebastien2012score, chiu2012study,nakamura2014merged}. Sebastien et al.~\cite{sebastien2012score} propose a list of different instrument-agnostic descriptors for difficulty classification.
The list of descriptors was further extended by Chen et al.~\cite{chiu2012study} proposing different feature spaces for measuring difficulty in the piano repertoire. Although a pitfall is that neither approach is reproducible without the code or datasets available, the main drawback is that they use instrument-agnostic attributes. In contrast, in the present research, we explore features related to the piano playing technique.  
Another work targeting piano difficulty is proposed by Nakamura et al.~\cite{nakamura2014merged,Nakamura2020} and deals with fingering frequencies and playing rate. The rationale for this proposal is that piano fingerings which occur less often, lead to increased difficulty. This method is extended to other tasks such as polyphonic transcription~\cite{nakamura2018towards}, rhythm transcription~\cite{nakamura2017rhythm} or score reductions~\cite{nakamura2015automatic,nakamura2018statistical}, making clear the importance of piano technique in the creation of music technology systems.

In this research, we aim at exploring several drawbacks of this approach~\cite{nakamura2014merged}, while looking at piano fingering as a proxy for difficulty. In contrast to their work which uses a concert-oriented dataset, we use a pedagogically motivated dataset. In addition, we classify overall difficulty and not solely instantaneous difficulty. 
Furthermore, we propose a thorough evaluation which is lacking in the existing literature~\cite{nakamura2014merged}. 
Nevertheless, we also use the Nakamura et al. approach~\cite{nakamura2014merged} to derive piano technique related features used as input for the machine learning models.

\section{Mikrokosmos-difficulty dataset}
\label{sec:dataset}

Béla Bartók's \textit{Mikrokosmos} Sz.\ 107 is a collection of 153 piano pieces published in six volumes between 1926 and 1939 and progressively ranked in terms of difficulty. 
Individual compositions go from extremely simple and easy beginner exercises to challenging advanced technical musical works.

\vspace{-0.2cm}
\begin{table}[h]
\begin{tabular}{lccc}
                                         & \multicolumn{1}{c}{Beginner} & \multicolumn{1}{c}{Moderate} & \multicolumn{1}{c}{Professional} \\ 
\multicolumn{1}{l}{n scores}   &  \pbox{3cm}{\vspace{2mm}62\vspace{2mm}}                                & \pbox{3cm}{\vspace{2mm} 54 \vspace{2mm}}                                  & \pbox{3cm}{\vspace{2mm} 31 \vspace{2mm}}                                      \\ 
\multicolumn{1}{l}{n notes}    & \pbox{3cm}{ \vspace{2mm} $\mu = 108.58$ \\ $\sigma = 57.16 $ \vspace{2mm}} & \pbox{3cm}{ \vspace{2mm} $\mu = 260.40$ \\ $\sigma = 111.00$ \vspace{2mm}} & \pbox{3cm}{ \vspace{2mm} $\mu = 650.06$ \\ $\sigma = 322.15$ \vspace{2mm}}     \\ 
\multicolumn{1}{l}{n bars} & \pbox{3cm}{ \vspace{2mm} $\mu = 20.37$ \\ $\sigma = 9.64$ \vspace{2mm}}    & \pbox{3cm}{ \vspace{2mm} $\mu = 33.35$ \\ $\sigma = 13.27$ \vspace{2mm}}   & \pbox{3cm}{ \vspace{2mm} $\mu = 63.12$ \\ $\sigma = 29.30$ \vspace{2mm}}       \\ 
\multicolumn{1}{l}{tempo}              & \pbox{3cm}{ \vspace{2mm} $\mu = 107.25$ \\ $\sigma = 24.48$ \vspace{2mm}}  & \pbox{3cm}{ \vspace{2mm} $\mu = 112.62$ \\ $\sigma = 59.75$ \vspace{2mm}}  & \pbox{3cm}{ \vspace{2mm} $\mu = 182.45$ \\ $\sigma = 113.13 $ \\}    \\ 
\end{tabular}
\caption{Means $\mu$ and the standard deviations $\sigma$ for the number of scores (n scores), notes (n notes), bars (n bars) and  tempo across the three levels of difficulty in the Mikrokosmos dataset.}
\label{tab:table_dataset}
\end{table}
\vspace{-0.2cm}

We propose the Mikrokosmos-difficulty dataset based on a subset of 146 pieces from the Mikrokosmos collection, discarding the 7 pieces composed for four hands performance. 
We group the pieces on three levels of difficulty according to the original classification in the the editions by the publishers Wiener and Henle Verlag. Correspondingly, the beginner level contains the pieces 1-66 (Volumes I and II), the moderate level contains the pieces 67-121 (Volumes III and IV), the professional level contains the pieces 122-153 (Volumes V and VI). 
%
%
%
The 6-volume-long collection is obtained from IMSLP in pdf format. The conversion to machine-readable symbolic representations in musicXML was done semi-automatically using a commercial Optical Music Recognition (OMR) software~\cite{photoscore}, and manual corrections. The statistics are presented in Table \ref{tab:table_dataset}, where we can see that Mikrokosmos-difficulty dataset is biased in lengths and tempo. Therefore, analyzing the technique difficulty in this dataset requires robustness to variations in length and tempo.


\section{Methodology}
\label{sec:methods}

We present five piano technique-based feature representations to analyse the performance difficulty and an additional baseline using solely the notes. Moreover, we use these features as input for two machine learning classifiers. The two methods are interpretable, and we may derive important feedback for music education applications.

\subsection{Feature representation of the piano technique}
\label{representation}

Automatic piano fingering systems aim to describe the  movements of hands  and fingers on the piano departing from the score. This task is related to piano technique and a proxy to modelling the difficulty of playing a score.
The current approaches in piano fingering go from expert systems~\cite{Parncutt1997} to local search algorithms~\cite{balliauw2017variable,pianoplayer} and, more recently, data-driven methods~\cite{nakamura2014merged,Nakamura2020}. 
Towards modelling difficulty, we derive piano technique features from two piano fingering approaches, a knowledge-driven system, Pianoplayer~\cite{pianoplayer}, and a data-driven system proposed by Nakamura et al.~\cite{Nakamura2020}. 

\begin{figure}[h!]
 \centering
 \includegraphics[width=72mm]{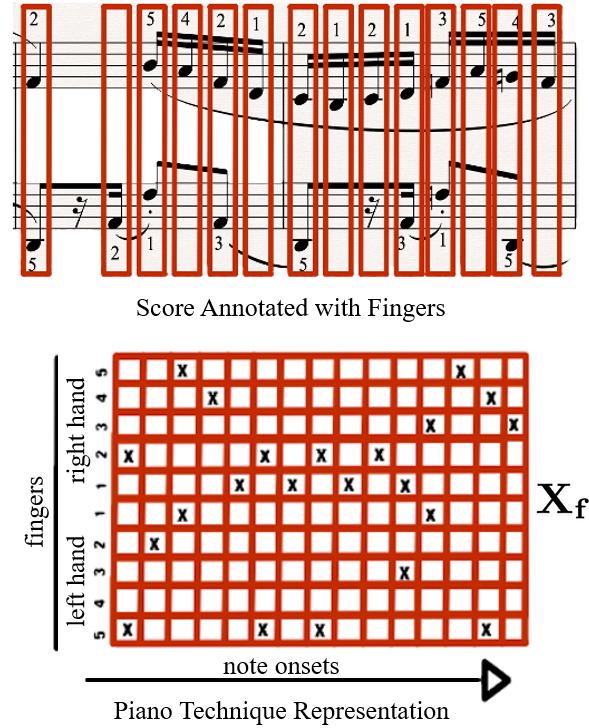}
 \caption{An example of piano fingering computed for a score (top) with Pianoplayer and Nakamura et al.~\cite{Nakamura2020} for which we derive the features \glpair{P}{F}, \glpair{P}{V}, \glpair{N}{F}, \glpair{N}{P}, and then the corresponding matrices $\matr{X_f}$}(bottom).
 \label{fig:fingers_representation}
\end{figure}

Pianoplayer~\cite{pianoplayer} is a local search algorithm grounded in dynamic programming and combinatorial optimisation, openly available under an MIT license. It computes which finger is playing each note by considering the following nine notes and the previous positions of the fingers. Pianoplayer carries out the optimisation according to hard-coded constraints and a cost function related to the finger velocity. From the cost function, we derive two features associated with each note: the finger which plays the note, \glpair{P}{F}, and the velocity associated with the finger, \glpair{P}{V}.



Nakamura et al.~\cite{Nakamura2020} train a Hidden Markov Model (HMM) on a 150 scores dataset annotated by more than four professional pianists -- data, code and models are available for noncommercial activities. 
In contrast to Pianoplayer which has hard-coded rules, this model learns implicitly the transition probabilities that model finger movements. 
We use the transition probability given by the HMM model to derive two features for a given note: the finger which plays the note, \glpair{N}{F}, and the transition probability associated to the note, \glpair{N}{P}.


Given a piano score and the features associated with each note \glpair{P}{F}, \glpair{P}{V}, \glpair{N}{F}, \glpair{N}{P}, we then form the matrices $\matr{X_f}(i,j)$ containing all the notes for each of the four features, where $i=1\ldots I$ and $I$ is the total number of note onsets in the piece, and $j=1\ldots 10$ are the fingers. In order to reduce the size of $\matr{X_f}$ and the computational complexity, we do not consider the note duration. 
In Figure~\ref{fig:fingers_representation} we present an example of how these matrices are formed. 
Note that the matrices $\matr{X_f}$ associated with \glpair{P}{F} and \glpair{N}{F} are binary matrices containing $1$ if the finger $j$ plays at note onset $i$, while the matrices $\matr{X_f}$ associated with \glpair{P}{V} and \glpair{N}{P} contain either the velocities or the transition probabilities associated with each note onset. 
As a baseline, we include a simple note onset representation \gl{K} as a feature agnostic to piano technique. We form the matrix $\matr{X_n}(i,k)$ containing $1$ if a note $k$ is played at the onset $i$ or $0$ otherwise, where $k=1\ldots 88$ are the notes that can be played on the piano.  

%

\vspace{-0.1cm}

\subsection{Classification methods}
\label{classiffication_methods}

We use the five features introduced in Section \ref{representation} as input for two machine learning methods, gradient-boosted trees and a GRU neural network with an attention mechanism. To that extent, we decided to use machine learning methods that offer some form of interpretability. This property may potentially help with understanding the difficult parts of a score. Correspondingly, the decision trees give us comprehensive explanations while the attention mechanism in the neural network points out which parts the network focuses on.


\noindent\textbf{Gradient-boosted trees -- \emph{XGBoost}}. 
We perform a two-step classification of a score into the three difficulty classes using a decision tree classifier, gradient boosted trees~\cite{gomez2017study}. 
Since this method works solely on fixed-sized inputs, we split the input feature matrices $\matr{X_f}$ and $\matr{X_n}$ into windows of size $w$, representing short-term score segments. The resulting matrices $\matr{\hat{X}_f}$ of size $(w\times 10)$ and $\matr{\hat{X}_n}$ of size $(w\times 88)$ are used as input to the classifier. Next, the predictions corresponding to each segment are averaged across the whole score to obtain a global difficulty.
%
%
Note that besides being interpretable, this method is more robust to noisy labels \cite{gomez2017study}.
In our case, we deal with a weakly supervised scenario because we have annotations solely for the whole piece, and not for each segment. Moreover, the difficulty of the segments may vary across a piece.


\noindent\textbf{GRU network with attention mechanism -- \emph{DeepGRU}}. 
Towards modelling time dependencies in the score and dealing with variable size scores as inputs, we use a recurrent neural network classifier. 
We modify an existing architecture used in multivariate time series classification, \emph{DeepGRU}~\cite{maghoumi2019deepgru}. The model lies in a set of stacked gated recurrent units (GRU), two fully connected (FC) layers and a global attention model.
The final two FC layers, using ReLU activations, take the output of the attention module and use a softmax classifier to create the probability distribution of the class labels. 
The GRUs stacked layers are able to model the time dependencies while the attention mechanism selectively attends to specific note onsets which are more important in the difficulty decision. We use the attention layer to visualise and understand the important notes. 


\section{Experiments}
\label{sec:experiments}

We evaluate the classification (balanced) accuracy of the machine learning models \emph{XGBoost} and \emph{DeepGRU} trained with proposed features \glpair{P}{F}, \glpair{P}{V}, \glpair{N}{F}, \glpair{N}{P}, \gl{K} on the Mikrokosmos-difficulty dataset we introduce in Section \ref{sec:dataset}. 
Towards assessing the impact of weak labels, we compare the 2-step classification, \emph{XGBoost (avg)} with a single step classification, where the predictions are not averaged across all segments, \emph{XGBoost (window)}.
In addition, we rank the scores using the output probability of the classifiers. The ranking is calculated by multiplying each class probability by their class number in the last layer.
We then compare this ranking with the original one proposed by Bartók and the one by the publisher Henle using Spearman's rank correlation.

\subsection{Experimental setup}

Because we want to keep the window size consistent with the $9$ notes used as a temporal context by Pianoplayer, we use segments of window size $w=9$. Hence, the input matrices are of size $9\times 10$ in the case of $\matr{\hat{X}_f}$ and $9\times 88$ in the case of $\matr{\hat{X}_n}$. Similarly, we set the stride $s=1$ in order to generate the maximum overlap between the windows. 
Note that we perform an ablation study to evaluate the effect of the window size. 

We use the gradient-boosted trees implementation in the XGBoost library~\cite{chen2016xgboost} for training the \emph{XGBoost (window)}. We use a random search over the training set (5-fold cross-validation) to tune seven hyper-parameters. In this case, we pick the parameters corresponding to the model with the best balanced accuracy on the validation set. 
The DeepGRU model is trained with the original parameters~\cite{maghoumi2019deepgru}: 20 epochs, an Adam optimizer with a learning rate of 0.002, a batch-size of 64 samples and using negative log likelihood loss as the criterion.

In both classification methods, the 80\% of data is used for training and 20\% for testing. We repeat the experiments for $50$ different initial random seeds which control the initialisation of the machine learning models and the train and test splits. We report means and standard deviations for the considered metrics across the $50$ seeds.



\subsection{Results}
\label{results}

The effect of different feature representations on the difficulty classification accuracy is presented in Table~\ref{tab:representation_comparison}. 
The results have to be interpreted considering the large standard deviations across the 50 seeds, which the small size of our dataset may cause.

\begin{table}[h!]
\small

\begin{center}

\caption{3-class classification accuracy (\%) for different features and machine learning models.}
\label{tab:representation_comparison}

\resizebox{\columnwidth}{!}{%
\begin{tabular}{lllllllll}
     
& \multicolumn{2}{l}{\emph{XGBoost (window)}} & \multicolumn{2}{l}{\emph{XGBoost (avg)}} & \multicolumn{2}{l}{\emph{DeepGRU}}\\ \cline{2-7} 
& train        & test    & train        & test         & train        & test  \\ 
\hline
$\gl{K}$              & $77\pm9$    & $51\pm7$ & $90\pm8$    & $65\pm10$    & $93\pm6$   & $64\pm8$   \\

$\glpair{N}{F}$ & $71\pm5$    & $51\pm5$ & $83\pm5$   & $62\pm8$   & $85\pm5$   & $72\pm8$   \\
$\glpair{P}{F}$           & $74\pm7$    & $52\pm4$ & $86\pm6$   & $64\pm8$   & $87\pm4$   & $73\pm6$   \\
$\glpair{N}{P}$             & $80\pm6$    & $57\pm5$ & $92\pm6$   & $67\pm8$   & $83\pm5$   & $71\pm8$   \\
$\glpair{P}{V}$         & $88\pm5$    & $62\pm6$ & $96\pm4$   & $68\pm10$    & $80\pm5$    & $\mathbf{78\pm6}$  
\end{tabular}
}
\end{center}
\end{table}

We observe that \emph{XGBoost (avg)} yields better results than \emph{XGBoost (window)}. Thus, averaging the results across the whole piece is better than considering stand-alone segments. 
%
We see that using transition probabilities, \glpair{N}{P}, and velocity information, \glpair{P}{V}, increases the performance compared to using the fingers solely, \glpair{P}{F}, \glpair{N}{F}. 
Comparing the two finger modelling methods, we see that Pianoplayer fingerings, \glpair{P}{}, perform better than the Nakamura fingerings, \glpair{N}{}. 
To our best knowledge, it is the first time Pianoplayer is explored in a research paper, and the results demonstrate its performance is comparable with the established Nakamura’s algorithm. 

The importance of finger modelling in piano difficulty analysis may be seen in the drastic improvement for the \emph{DeepGRU} case, where all the fingering strategies improved the classification accuracy by a large margin compared to features generated directly from sheet music without finger modelling ($\gl{K}$).

The ranking performance using Bartók and Henle difficulty rankings is presented in Table \ref{tab:ranking_comparison}. Even though the models are trained using solely 3-class labels, the difficulty rankings we derive evaluated against the rankings provided by the composer (Bartók) and the publisher (Henle) achieve high correlation. 
%

\vspace{-0.1cm}
\begin{table}[h!]
\small
\begin{center}

\caption{Spearman's rank correlation coefficient between the rankings derived from the output probabilities of the classifiers and the Bartók and Henle rankings}
\label{tab:ranking_comparison}
\begin{tabular}{lllllll}

& \multicolumn{2}{l}{\emph{XGBoost (avg)}} & \multicolumn{2}{l}{\emph{DeepGRU}} \\ \cline{2-5} 
& Bartók        & Henle         & Bartók        & Henle  \\ 
\hline
$\gl{K}$  & $.74\pm.09$ & $.75\pm.08$ & $.71\pm.08$ & $.71\pm.09$ \\
$\glpair{N}{F}$ & $.70\pm.08$ & $.66\pm.08$ & $.82\pm.06$ & $.81\pm.06$ \\
$\glpair{P}{F}$ & $.69\pm.09$ & $.64\pm.11$ & $.80\pm.06$ & $.80\pm.06$ \\
$\glpair{N}{P}$  & $.78\pm.07$ & $.73\pm.09$ & $.84\pm.06$ & $.83\pm.06$ \\
$\glpair{P}{V}$ & $.81\pm.08$ & $.78\pm.10$ & $\mathbf{.86\pm.05}$ & $\mathbf{.86\pm.05}$ 
\end{tabular}
\end{center}
\end{table}
\vspace{-0.2cm}

The ablation study of the window size for the \emph{XGBoost} method is shown in Table~\ref{tab:windowsize_comparison}. Considering the high standard deviation, we concluded that the change in window size is not statistically significant.  For this reason, we decided to stick with the window size of the Pianoplayer method, which is $w=9$.



\begin{table}[h!]
\begin{center}
\caption{Evaluation considering different window size for \emph{XGBoost (avg)}. Results are presented in terms of 3-class classification accuracy (\%), Spearman rank correlation coefficient for Bartók and Henle.}
\label{tab:windowsize_comparison}
\begin{tabular}{llll}
window size     & 3-class acc     & Bartók rank        & Henle rank          \\ \hline 
$w=1$  & $69\pm12$ & $.81\pm.08$ & $.77\pm.10$ \\
$w=3$  & $69\pm10$ & $.81\pm.08$ & $.78\pm.10$ \\
$w=5$  & $69\pm10$ & $.81\pm.09$ & $.78\pm.10$ \\
$w=9$  & $68\pm10$ & $.81\pm.08$ & $.78\pm.10$ \\
$w=13$ & $69\pm10$ & $\mathbf{.83\pm.07}$ & $\mathbf{.82\pm.07}$  \\
$w=19$ & $\mathbf{70\pm9}$  & $.82\pm.06$ & $.80\pm.07$ 
\end{tabular}
\end{center}
\end{table}
\vspace{-0.5cm}

\subsection{Local Difficulty Feedback}

As a practical application to music learning, giving feedback related to the local difficulty of a piece allows students and teachers to focus and improve on the most difficult passages.
Because \emph{XGBoost} and \emph{DeepGRU} are interpretable, we derive local difficulty feedback for both of the methods. 
We created an online tool to explore the two different types of feedback for the Mikrokosmos-difficulty dataset~\footnote{At: \url{https://musiccritic.upf.edu/mikrokosmos}}.

\begin{figure}[h!]
 \centering
 \includegraphics[width=80mm]{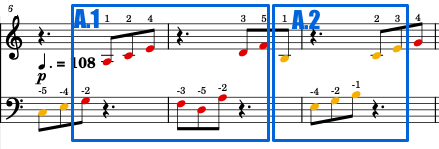}
 \caption{Mikrokosmos Sz.\ 107 no.\ 142. Béla Bartók excerpt. Local difficulty feedback from \emph{XGBoost (window)} with \glpair{P}{V} as input feature. The score is from the test set of a random seed.}
 \label{fig:feedback_2steps}
\end{figure}

\noindent\textbf{Window-based feedback.} 
The difficulty of the windows pertaining to a score may vary considerably. In the case of the 2-step \emph{XGBoost (window)} classification, we obtain an output probability for each window. Considering the maximum overlapping for the windows for a stride of $s=1$, we derive a probability for each note onset. Therefore, this feedback is an indicator of the local difficulty corresponding to that note onset. 
%
An example of visualisation of the local difficulty may be seen in Figure~\ref{fig:feedback_2steps}. 
The difficulty levels in the Mikrokosmos-difficulty dataset are displayed over the score with three different colours
We colour level 1 notes with green, level 2 notes with yellow and level 3 with red. 
%
In the excerpt shown in Figure~\ref{fig:feedback_2steps} we note how the difficulty changes within the same piece. 
We observe that the part we highlight as A.1 has asymmetric changes between the two hands while the A.2 part is a simple arpeggio. Consequently, A.1 is marked in red while A.2 is yellow.

\begin{figure}[h!]
 \centering
 \includegraphics[width=90mm]{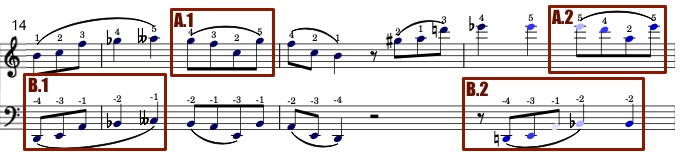}
 \caption{Mikrokosmos Sz.\ 107 no.\ 109. Béla Bartók excerpt. Local difficulty feedback from \emph{DeepGRU} approach with \glpair{P}{V} as input feature. The score is from the test set of a random seed. }
 \label{fig:feedback_attention}
\end{figure}

\noindent\textbf{Attention-based feedback.} This method grounds on the weights of the attention layer in the \emph{DeepGRU} model. Attention tells us the notes on which the model is focusing. 
The most interesting aspect of the present feedback is dealing with repetitions that occur in the short term. We can see a particular example in Figure~\ref{fig:feedback_attention}.In this plot the attention weights corresponding to each note control the intensity of the color.  
The pairs A.1-A.2 and B.1-B.2 are performed with the same physical movement. Therefore, in the second pattern, A.2 and B.2 ,respectively, the method pays less attention. The attention reasoning is similar to how music students should study. When dealing with two similar patterns, students should study the first one very well and repeat the second one by imitating the first one.

\vspace{-0.3cm}

\section{Conclusions and future work}
\label{sec:conclusion}
In this work, we proposed a methodology to classify piano scores based on piano technique features. Namely, we derive piano fingering features from two approaches~\cite{Nakamura2020,pianoplayer}, and a simple baseline based on using solely the notes in the score. 
Towards training and testing machine learning methods for this task, we distributed the Mikrokosmos-difficulty dataset.
We evaluated two interpretable machine learning methods on the proposed dataset to show that piano technique features are good predictors for the difficulty of performing a piece. 
In addition to the openly available source code, models and data, we also provide an online demo which visualizes the segmental difficulty, which is crucial in educational feedback. We acknowledge that the small size of the dataset may lead to over-fitting biases and large standard deviations for the bootstrapped experiments. In future, we plan to extend this work with larger score dataset with difficulty annotations and other instruments.

%
%
%

\bibliographystyle{IEEEbib}
\bibliography{main}

\end{document}